\magnification\magstep2

\def\apres{\vskip 0.5 true cm}
\def\avant{\vskip 0.8 true cm}
% \font\itdroit=cmub10
% \font\unhb=cmdunhb10
% \font\xsl=cmbxsl10
\nopagenumbers
% \font\msym=msym10
\overfullrule=0pt

% \font\msim=msym10
\def\gr{\bf R}

\def\grn{\bf N}

\def\grc{\bf C}

\rightline{Published in:}
\rightline{Phys.~Lett.~A}
\rightline{206 (1995) 13-17}

\vskip 2 true cm

\noindent
{\bf Statistical mechanics of $qp$-bosons in $D$ dimensions}

\vskip 2.3 true cm

\noindent
M. Daoud and M. Kibler

\medskip
\baselineskip = 0.5 true cm

\noindent
{\it Institut de Physique Nucliaire de Lyon,

\noindent
IN2P3-CNRS et Universiti Claude Bernard,

\noindent
43 Bd.~du 11 Novembre 1918, F-69622 Villeurbanne Cedex, France}

\vskip 3 true cm

This paper is concerned with statistical properties of a
gas of $qp$-bosons without interaction. Some thermodynamical
functions for such a system in $D$ dimensions are derived.
Bose-Einstein condensation is discussed in terms of the
parameters $q$ and $p$. Finally, the second-order correlation
function of a gas of photons is calculated.

\vfill\eject
\baselineskip = 0.8 true cm

1. In the last twenty years, different types of deformed bosons have been
       introduced [1-8].
        There use for obtaining realisations of quantum algebras
       is now widely widespread (e.g., see [4-6,8]). We shall deal here with
       $qp$-bosons [6-8] associated to the annihilation ($a$),
       creation ($a^+$) and number operator ($N$)
       characterized through their action
$$
a   | n \rangle = \sqrt {[[n  ]]_{qp}} | n-1 \rangle, \quad
a^+ | n \rangle = \sqrt {[[n+1]]_{qp}} | n+1 \rangle, \quad
N   | n \rangle = n | n \rangle
\eqno (1)
$$
on the ordinary Fock space
${\cal F} := \{ | n \rangle \ : \ n \in \grn \}$.
(The operators $a$, $a^+$ and $N$ span a $qp$-deformed Weyl-Heisenberg
algebra.)
The $qp$-deformed number in
       (1) are defined by
$$
[[x]]_{qp} := {{q^x - p^x} \over {q-p}}.
\eqno (2)
$$
The particular cases $p=1$ and $p=q^{-1}$ correspond to $q$-deformations
mainly used in the mathematical and physical literature, respectively. In
the general case, Hermitean conjugation requirements demand that
$(q,p) \in \gr  \times \gr $, or
$(q,p) \in \grc \times \grc$ with $p = \bar q         $, or
$(q,p) \in S^1  \times S^1 $ with $p = \bar q =q^{-1}$ [9].

Numerous works have been devoted to the study of the statistical
thermodynamical properties of
a gas of deformed bosons (mainly $q$-bosons) without interaction [10-21].
(When writing this Letter, we became aware of a paper [21] dealing with
$qp$-bosons.) In particular, the phenomenon of Bose-Einstein condensation of
deformed bosons has been studied by several authors [10,13,14,16,20,21].
Furthermore,
the deformation of the  correlation function of order two  for a gas of photons
has been recently investigated by Man'ko {\it et al.}~[15]. Most of the
preceding works concern $q$- and $qp$-bosons in $D=2$ or $3$ dimensions
and are restricted to $q \in \gr$ [10-20] and
                  $(q,p) \in \gr \times \gr$ [21].

We derive in this Letter the (thermodynamical)
properties of a gas of $qp$-bosons in $D$ dimensions for parameters $q$ and $p$
varying in five domains. The Bose-Einstein condensation of
$qp$-bosons is discussed in a new approach. Finally, we obtain
the correlation function $g^{(2)}$
for a gas of photons described
by $qp$-bosons.

\avant

2. We start with the Hamiltonian (in a second-quantized form)
$$
H := \sum_k H_k, \quad H_k := (E_k - \mu) N_k,
\eqno (3)
$$
which describes a gas of $qp$-bosons without interaction. In eq.~(3),
$\mu$ is the chemical potential while $E_k$ and $N_k$ are the kinetic
energy of a $qp$-boson and the number operator for the $qp$-bosons,
 in the $k$ mode, respectively. The $qp$-analogue of the Bose factor
for the $k$ mode
is then
$$
(f_k)_{qp} := {1\over Z} {\rm tr}
                                 \big ( {\rm e}^{-\beta H} a^+_k a_k \big ),
\eqno (4)
$$
where $Z := {\rm tr} \left( {\rm e}^{-\beta H} \right)$
is the partition function
and $\beta = (k_BT)^{-1}$ the reciprocal temperature. As a trivial
result, we have
$$
Z = \prod_{k} { 1 \over 1 - {\rm e}^{-\eta} }, \quad \eta := \beta (E_k - \mu).
\eqno (5)
$$
The partition function $Z$ is thus independent of the
deformation parameters $q$ and $p$.
Then, the Bose factor $(f_k)_{qp}$ reads
$$
(f_k)_{qp} =
(1 - {\rm e}^{-\eta}) {\rm tr}
                               \big ( {\rm e}^{-\eta N_k} [[N_k]]_{qp} \big ),
\eqno (6{\rm a})
$$
which can be shown to converge in each of the following five cases~:
  (i) $(q,p) \in \gr^+ \times \gr^+$ with
$0 < q < {\rm e}^{-\beta \mu}$ and
$0 < p < {\rm e}^{-\beta \mu}$;
 (ii) $(q,p) \in \gr^+ \times \gr^+$ with
$p=q^{-1}$ and
${\rm e}^{ \beta \mu} < q <
 {\rm e}^{-\beta \mu}$;
(iii) $(q,p) \in \gr^+ \times \gr^+$ with
$p=1$ and
$0 < q < {\rm e}^{-\beta \mu}$;
(iv) $(q,p) \in \grc \times \grc$ with
$p=\bar q$ and
$0 < |q| <
 {\rm e}^{-\beta \mu}$;
(v) $(q,p) \in S^1 \times S^1$ with
$p=\bar q = q^{-1}$.

At this stage, it is important to justify the choices made for
eqs.~(3) and (4). The $qp$-bosons are not new particles. They
are still bosons. Thus, (3) and (4) correspond to the basic
formulas for the statistical mechanics of bosons. Here, we do
not deform the basic formulas. The trace and the exponential
in eq.~(4) are nondeformed functions. We only use (through
the formalism of second quantization) creation and annihilation
operators that satisfy $qp$-deformed commutation relations. The
form chosen for $H_k$ in eq.~(3) is nothing but the usual one
where the number operator takes its eigenvalues in $\grn$. The
deformation cannot enter the theory at this level by replacing
$N_k$ by $a_k^+a_k$ in (3) since we would obtain an ``occupation
number'' $a_k^+a_k = [[N_k]]_{qp}$ that would not have its eigenvalues
in $\grn$. On the contrary, we must keep $a_k^+a_k$ rather than
$N_k$ in (4). Indeed, should we use $N_k$ in place of $a_k^+a_k$ in
(4), we would get a nondeformed value of the Bose factor. As a conclusion,
eqs.~(3) and (4) constitute a reasonable starting point for a (minimal)
deformation of the Bose distribution.

Following ref.~[9], we can use the parametrisation
$$
q = {\rm e}^{\varphi \cos \tau} {\rm e}^{+\varphi \sin \tau}, \quad
p = {\rm e}^{\varphi \cos \tau} {\rm e}^{-\varphi \sin \tau}
\eqno (7)
$$
for $(q,p) \in \gr   \times \gr  $ and the parametrisation
$$
q = {\rm e}^{\varphi \cos \tau} {\rm e}^{+ {\rm i}\varphi \sin \tau}, \quad
p = {\rm e}^{\varphi \cos \tau} {\rm e}^{- {\rm i}\varphi \sin \tau}
\eqno (8)
$$
for $(q,p) \in \grc \times \grc$ or
    $(q,p) \in S^1  \times S^1 $, where $\varphi \in \gr^*$ and
$ - \pi/2 \le \tau
\le \pi/2$,
so that the limiting situation $q=p=1$ corresponds to
$\varphi \to 0$.

The obtained expression for $(f_k)_{qp}$ is
$$
(f_k)_{qp} =
{q-1\over q-p} {1\over {\rm e}^\eta - q} +
{p-1\over p-q} {1\over {\rm e}^\eta - p}
\eqno (9{\rm a})
$$
or alternatively
$$
(f_k)_{qp} = { {\rm e}^{\eta} - 1 \over
              ({\rm e}^{\eta} - q)
              ({\rm e}^{\eta} - p) },
\eqno (10)
$$
which clearly exhibits the $q \leftrightarrow p$ symmetry.
In the (five) limiting situations where $\varphi \to 0$,
we recover the ordinary Bose distribution
$f_k = ({\rm e}^\eta -1)^{-1}$.
Note that $(f_k)_{qp}$ can be rewritten as
$$
(f_k)_{qp} = {q-1\over q-p} {f_k \over 1 + (1-q)f_k}
           + {p-1\over p-q} {f_k \over 1 + (1-p)f_k}
\eqno (9{\rm b})
$$
in term of $f_k$.

For the purpose of practical applications (in Section 3), it is useful
to know the development of the distribution $(f_k)_{qp}$ in integer
series. In this respect, we have
$$
(f_k)_{qp} =
\sum^\infty_{j=0} {\rm e}^{ - \eta (j+1) } \left ( {q-1\over q-p} q^j +
                                                   {p-1\over p-q} p^j \right ),
\eqno (6{\rm b})
$$
which reduces to the expansion $\sum^{\infty}_{j=1} {\rm e}^{ -\eta j }$ of the
ordinary Bose factor when $\varphi \to 0$.

\avant

3. In the thermodynamical limit, the energy spectrum of the system of
$qp$-bosons may be considered as a continuum. Thus, $(f_k)_{qp}$ is
replaced by
the $qp$-dependent factor $f(\epsilon)$ which is $(f_k)_{qp}$
with $E_k = \varepsilon$. Therefore, in physical applications,
we have to consider integrals of the type
$$
J_s := \int^\infty_0 \varepsilon^s f(\varepsilon) d\varepsilon.
\eqno (11)
$$
By using the development (6b), eq.~(11) yields
$$
J_s = \Gamma (s+1) (k_B T)^{s+1} \sigma (s+1)_{qp}, \quad s>-1,
\eqno (12)
$$
where
$$
\sigma(s+1)_{qp} = \sum^\infty_{j=0}
{{\rm e}^{ \beta \mu (j+1)} \over (j+1)^{s+1}}
\big ( [[j+1]]_{qp} - [[j]]_{qp} \big ).
\eqno (13)
$$
[In eq.~(12), $\Gamma$ is the Euler integral of the
second type.] In the case where $\mu = 0$, we have
$\sigma (s+1)_{qp} \rightarrow \zeta (s+1)$ for $\varphi \to 0$,
where $\zeta(s+1)$ is the Riemann Zeta series (that converges for $s>0$).

We are now in a position to derive some thermodynamical quantities, in
$D$ dimensions, for a free gas of $N(D)$ $qp$-bosons contained in
a volume $V(D)$. The density $\rho(D) = N(D)/V(D)$ is given by [14]
$$
\rho (D)
= N_0(D) \Gamma ({D \over 2}) (k_B T)^{{D \over 2}} \sigma ({D \over 2})_{qp},
\eqno (14)
$$
where
$$
N_0(D) :=
{ 1 \over 2 (2 \pi)^{D \over 2} }
{ D \over {\Gamma({D \over 2} + 1)} }
g {m^{D \over 2} \over \hbar^D}.
\eqno (15)
$$
In eq.~(15), $g$ is the degree of (spin) degeneracy and $m$ the mass of a
$qp$-boson. The total energy can then be calculated to be
$$
E(D) = N_0(D) V(D)               \Gamma \big ({D \over 2} + 1 \big )
       (k_B T)^{{D \over 2} + 1} \sigma \big ({D \over 2} + 1 \big )_{qp}.
\eqno (16)
$$
The specific heat at constant volume easily follows from
$C_V = \big ( {\partial E \over \partial T } \big )_V$. We obtain
$$
C_V(D) = {D \over 2} N(D) k_B
\bigg [
       \big  ( {D \over 2} + 1 \big )
{\sigma \big ( {D \over 2} + 1  \big )_{qp} \over
 \sigma \big ( {D \over 2}      \big )_{qp}}
             - {D \over 2}
{\sigma \big ( {D \over 2}      \big )_{qp} \over
 \sigma \big ( {D \over 2} - 1  \big )_{qp}}
\bigg ].
\eqno (17)
$$
Furthermore, the entropy is
$$
S(D) = N(D) k_B
\bigg [
        \big ( {D \over 2} + 1 \big )
{\sigma \big ( {D \over 2} + 1 \big )_{qp} \over
 \sigma \big ( {D \over 2}     \big )_{qp}} - {\mu \over k_B T}
\bigg ].
\eqno (18)
$$
Note that the state equation for the gas of $qp$-bosons is
$$
p(D) = {2 \over D} { E(D) \over V(D) },
\eqno (19)
$$
so that the pressure $p(D)$ assumes the same form as in the nondeformed case.

Finally, other thermodynamical quantities may be determined
in a simple manner  by  using  the  thermodynamic potential
$$
\Omega(D) = - {2 \over D} N_0(D) V(D) J_{{D \over 2}}
          = - {2 \over D} E(D),
\eqno (20)
$$
where $J_{{D\over 2}}$ is given by (12) and (13).

\avant

4. We now examine the condensation of a system
   of $qp$-bosons in $D$ dimensions.
The corresponding density for such a system
is  given  by (14) which can be
rewritten as
$$
\rho (D) = N_0 (D) J_{ {D \over 2} - 1 }
\eqno (21)
$$
in function of the integral $J_{ {D \over 2} - 1 }$.
As in the case of ordinary bosons, we define the Bose temperature $T_B(D)$
for the $qp$-bosons by taking $\mu = 0$. By introducing $\mu = 0$ in
$J_{ {D \over 2} - 1}$,
we find that the temperature below which we obtain
a Bose-Einstein condensation is given by
$$
T_B(D) =
{1 \over k_B}
\bigg [ {\rho(D) \over N_0(D)}
{1 \over \Gamma ({D \over 2}) \sigma_0 ({D \over 2})_{qp}}
\bigg ]^{2 \over D},
\eqno (22)
$$
where
$$
\sigma_0({D \over 2})_{qp} = \sum^\infty_{j=0}
{1 \over (j+1)^{D \over 2}} \big ( [[j+1]]_{qp} - [[j]]_{qp} \big ).
\eqno (23)
$$
The study of the convergence of the series (23)
determines if Bose-Einstein condensation takes place or not.

In the limiting cases where $\varphi \to 0$, eqs.~(22) and (23) give back
the well-known results according to which there is (respectively, there is not)
condensation in $D = 3$ (respectively, $D = 2$) dimensions. As an illustration,
we recover that the Bose temperature is
$$
T_B(3) =
{(2 \pi) \hbar^2 \over k_B m}
\left [ {\rho(3) \over 2.612 g} \right ]^ {2 \over 3}
\eqno (24)
$$
in the case $D$ = 3.

Returning to the general case,  we  can  prove
the following results valid for $D$ arbitrary.

 {(i)} Case $(q,p) \in \gr  \times \gr $~: the series (23) converges for
$0 < q < 1$ and
$0 < p < 1$. Thus, there is Bose-Einstein condensation
of $qp$-bosons for $0 < q < 1$ and $0 < p < 1$. In addition, there is
no Bose-Einstein condensation (i.e., $T_B(D)=0$)
for $q$-bosons corresponding to $p=q^{-1}$.
In contradistinction, Bose-Einstein condensation
takes place for $q$-bosons corresponding to $p=1$ when $0< q <1$.

{(ii)} Case $(q,p) \in \grc \times \grc$~: Bose-Einstein condensation
occurs either for the $q$-bosons corresponding to $p=\bar q$ when
$0 < |q| < 1$
           or for all $q$-bosons corresponding to $p=\bar q= q^{-1}$.

These results have to be compared to the ones corresponding
to the limiting situation where $\varphi \to 0$ for which
Bose-Einstein condensation exists only when $D \geq 3$.

\avant

5. As a final facet of this Letter, we would like
to discuss the consequence of a $qp$-deformation
of the correlation function $g^{(2)}$ of order two associated to
the radiation field. In the nondeformed case, we know that $g^{(2)}$ takes
two values, viz., $g^{(2)} = 1$
for a coherent monomode radiation and $g^{(2)} = 2$
for a chaotic  monomode radiation. An interesting question arises~:
Is it possible to
interpolate between the latter two values by replacing the ordinary bosons of
the radiation field by $qp$-bosons~?

Our basic hypothesis is to describe the radiation field
by an assembly of $qp$-bosons with the Hamiltonian
$$
h = \sum_k h_k, \quad h_k = \hbar \omega_k N_k.
\eqno (25)
$$
The corresponding Bose statistical distribution for the $k$ mode is then
given by (10) where $\eta$ is replaced by $\xi = \beta \hbar \omega_k$.
In the case of $qp$-bosons, we adopt the definition
$$
g^{(2)} = {\langle a^+a^+aa \rangle \over \langle a^+ a \rangle^2},
\eqno (26)
$$
where $a$ and $a^+$ stand for the annihilation
and creation operators for the $k$ mode.
The abbreviation $\langle  X \rangle$ in (26) denotes the
mean statistical value $Z^{-1} {\rm tr} \left( {\rm e}^{-\beta h} X \right)$
for an operator $X$ defined on the Fock space ${\cal F}$.
Equation (26) can be developed as
$$
g^{(2)} =
p^{-1}
{\langle (a^+ a)^2 \rangle \over \langle a^+ a \rangle^2} -
(qp)^{-1}
{\langle q^N a^+a  \rangle \over \langle a^+ a \rangle^2}.
\eqno (27)
$$
Of course, $\langle a^+ a \rangle = (f_k)_{qp}$
as given by (10) with $\eta \equiv \xi$.
In addition, the other average values in (27) can be
calculated to be
$$
\langle (a^+ a)^2 \rangle =
{({\rm e}^\xi -   1) ({\rm e}^\xi + qp) \over
 ({\rm e}^\xi - q^2) ({\rm e}^\xi - qp)
 ({\rm e}^\xi - p^2)}
\eqno (28)
$$
and
$$
\langle q^N a^+ a \rangle =
q { {\rm e}^\xi - 1   \over
   ({\rm e}^\xi - q^2)
   ({\rm e}^\xi - qp )}.
\eqno (29)
$$
Finally, we obtain
$$
g^{(2)} = (q + p)
{1 \over {\rm e}^\xi - 1   }
       {({\rm e}^\xi - q  )^2
        ({\rm e}^\xi - p  )^2 \over
        ({\rm e}^\xi - q^2)
        ({\rm e}^\xi - qp )
        ({\rm e}^\xi - p^2)},
\eqno (30)
$$
with convergence conditions that parallel the ones for $(f_k)_{qp}$,
cf.~cases (i)-(v) in Section 1 where
$ - \beta \mu$ must be replaced by $\xi/2$.
[For instance, when $(q,p) \in \gr^+ \times \gr^+$, we must have
$0 < q < {\rm e}^{\xi \over 2}$ and
$0 < p < {\rm e}^{\xi \over 2}$.]
The $qp$-deformed factor $g^{(2)}$ depends on the parameters $q$
and $p$ in a symmetrical manner ($q \leftrightarrow p$ symmetry).
It also presents a dependence on the energy $\hbar \omega_k$
of the $k$ mode and on the temperature $T$.
In the limiting cases $\varphi \to 0$,
we get $g^{(2)} = 2$ that turns out to be the
value of the correlation function
for the choatic monomode radiation of the black body.
Let us now examine the cases of low temperatures and high energies.
In these cases,
${\rm e}^\xi$ is the dominating term in each of the differences
occuring in eq.~(30). Therefore, we have
$$
g^{(2)} \sim q + p
\eqno (31)
$$
at low temperature or high energy.
It is thus possible to reach the value $g^{(2)} = 1$
without employing coherent states.
Equation (31) shows that we can interpolate
in a continuous way from  $g^{(2)} = 1$ (coherent phase) to
                          $g^{(2)} = 2$ (chaotic  phase).
It should be observed that we can even obtain either
 $g^{(2)} > 2$ or
 $g^{(2)} < 1$ from eq.~(30). The situation where
 $g^{(2)} < 1$ may be interesting for describing
antibunching effects of the light field (arising from the
``nonclassical'' nature of the radiation field, cf.~ref.~[21]).

\avant

6. In this work, we concentrated on the influence of a $qp$-deformation of the
Weyl-Heisenberg algebra on the statistical properties of a gas of bosons in $D$
dimensions. We discussed the phenomenon of Bose-Einstein condensation in terms
of the deformation parameters $q$ and $p$. Bose-Einstein condensation occurs in
most of the domains where the $q$ and $p$ parameters may vary. In the case
$(q,p) \in \gr^+ \times \gr^+$ with $p = q^{-1}$ and $D$ arbitrary,
Bose-Einstein condensation
does not take place. For the latter case, we may ask the
question~: Are there any topological interactions (leading to some
quasi-fermionic behavior through some exclusion rule) that prevent the
condensation of the corresponding $q$-bosons~?

An important result of the present work concerns the correlation function
$g^{(2)}$ for a gas of photons. The $qp$-deformation considered here makes
it possible to obtain a continuum of values for
$g^{(2)}$. In particular, such values may be greater than 2 or lesser than 1.

To close this Letter, it should be observed that alternative choices,
non-linear in $N_k$, are possible for the Hamiltonians
    $H$ (eq.~(3))
and $h$ (eq.~(25)), choices that seem to reflect more deeply the quantum group
structure of the $qp$-deformed Weyl-Heisenberg algebra. In this direction,
we may mention, in the case $p=q^{-1}$, the works of
refs.~[11,12,15,19,20]. We have justified above
our choice for eq.~(3) (and eq.~(4)).
The case of eq.~(25) is more delicate. It
would be interesting to investigate the situation where
$h_k = (1/2) \hbar \omega_k ([[N_k]]_{qp} + [[N_k + 1]]_{qp})$
since the latter
expression corresponds to a $qp$-deformed oscillator. The
corresponding partition function can then be expressed in series of Bessel
functions with some convergence problem.
Our choice for eq.~(25) does not lead
to convergence problem and represents
the first attempt to derive a deformation
of the second-order correlation function
in a nonperturbative way. We hope to return
on these matters in the future.

\avant

\noindent {\bf References}

\apres

\noindent
\item{[1]}
M.~Arik and D.D.~Coon, J.~Math.~Phys.~17 (1976) 524.

\noindent
\item{[2]}
V.~Kuryshkin, Ann.~Fond.~Louis de Broglie 5 (1980) 111.

\noindent
\item{[3]}
A.~Jannussis, G.~Brodimas, D.~Sourlas, K.~Vlachos,
P.~Siafarikas and L.~Papaloucas, Had.~J.~6 (1983) 1653.

\noindent
\item{[4]}
A.J.~Macfarlane, J.~Phys.~A 22 (1989) 4581.

\noindent
\item{[5]}
L.C.~Biedenharn, J.~Phys.~A 22 (1989) L873.

\noindent
\item{[6]}
R.~Chakrabarti and R.~Jagannathan, J.~Phys.~A 24 (1991) L711.

\noindent
\item{[7]}
J.~Katriel and M.~Kibler, J.~Phys.~A 25 (1992) 2683.

\noindent
\item{[8]}
M.R.~Kibler, in Symmetry and Structural Properties
of Condensed Matter, eds. Florek, D. Lipi\'nski and T. Lulek
(World Scientific, Singapore, 1993).

\noindent
\item{[9]}
R. Barbier, J. Meyer and M. Kibler, Int. J. Mod. Phys. E (in press).

\noindent
\item{[10]}
C.R. Lee and J.-P. Yu, Phys.~Lett.~A 150 (1990) 63;
                                     164 (1992) 164.
See also~: C.-R. Lee,
Chinese J. Phys. 30 (1992) 491.

\noindent
\item{[11]}
M.-l. Ge and G. Su, J. Phys. A 24 (1991) L721.
% $q$-oscillateurs, fonction de Green doublement dipendante du temps
% $\Rightarrow$ fonction de distribution statistique des $q$-bosons
% et des $q$-fermions.
% ancien [12]

\noindent
\item{[12]}
M.A.~Mart\'\i n-Delgado, J.~Phys.~A 24 (1991) L1285.
% $q$-oscillateur, $q$-diformation de la distribution de Planck.
% ancien [11]

\noindent
\item{[13]}
G. Su and M.-l. Ge, Phys. Lett. A 173 (1993) 17.
% Condensation de Bose-Einstein.

\noindent
\item{[14]}
J.A. Tuszy\'nski, J.L. Rubin, J. Meyer and M. Kibler,
Phys. Lett. A 175 (1993) 173.
% Difinition de $f_k$. Condensation de B-E en $D=2$ et $D=3$.

\noindent
\item{[15]}
V.I. Man'ko, G. Marmo, S. Solimeno and F. Zaccaria,
Phys. Lett. A 176 (1993) 173.
% $q$-oscillateur et facteur $g^{(2)}$ de manihre perturbative.
% ancien [18]

\noindent
\item{[16]}
R.-R. Hsu and C.-R. Lee, Phys. Lett. A 180 (1993) 314.
% Mithode de la balance(?) $\Rightarrow$ fonction de distribution.

\noindent
\item{[17]}
S. Vokos and C. Zackos, ANL-HEP-CP-93-39.
% a \'et\'e publi\'e~?
% $q$-bosons ($H = \omega N$), calcul de $f_k$. Notre
% $(f_k)_{qp} \mathop{\hbox to 10 mm{\rightarrowfill}}\limits_{p=q^{-1}}
% f_k$ (Vokos et Zackos).
% ancien [19]

\noindent
\item{[18]}
M. Chaichian, R.G. Felipe and C. Montonen,
J. Phys. A 26 (1993) 4017.
% $q$-bosons $(H = \omega N)$, calcul de $f_k$. Notre
% $(f_k)_{qp} \mathop{\hbox to 10 mm{\rightarrowfill}}\limits_{q^{-1}=p}
% f_k$ (Chaichian {\it et al.}).
% ancien [15]

\noindent
\item{[19]}
R.K. Gupta, C.T. Bach and H. Rosu, J. Phys. A 27 (1994) 1427.
% $q$-diformation de la distribution de Planck ($q$ est complexe).
% ancien [17]

\noindent
\item{[20]}
M.A. R.-Monteiro, I. Roditi and L.M.C.S. Rodrigues,
Phys. Lett. A 188 (1994) 11.
% traite des q-bosons

\noindent
\item{[21]}
R.-S. Gong, Phys. Lett. A 199 (1995) 81.
% traite des qp-bosons
% ancien [20]

\noindent
\item{[22]}
D.F. Walls, Nature 306 (1983) 141.
% ancien [21]

\bye